\pdfoutput=1
\documentclass[aps,pre,groupedaddress,showpacs,showkeys,twocolumn,amsmath,amsfonts,amssymb,superscriptaddress]{revtex4-1}
\usepackage{graphicx}
\usepackage[utf8x]{inputenc}
\usepackage{hyperref}
\usepackage{xcolor}
\usepackage{bm}

\hypersetup{
    colorlinks,
    linkcolor={red!50!black},
    citecolor={blue!60!black},
    urlcolor={blue!50!black}
}

\graphicspath{{fig/}} 

\begin{document}


\title{Entanglement estimation in non-optimal qubit states}


\author{Stefano Scali}
\email[]{ss1116@exeter.ac.uk}
\affiliation{DSFTA, University of Siena, Via Roma 56, 53100 Siena, Italy}
\affiliation{Department of Physics, University of Cambridge, Cambridge CB3 0HE, United Kingdom}
\affiliation{Department of Physics and Astronomy, University of Exeter, Exeter EX4 4QL, United Kingdom}
\author{Roberto Franzosi}
\email[]{roberto.franzosi@ino.it}
\affiliation{QSTAR \& CNR - Istituto Nazionale di Ottica, Largo Enrico Fermi 2, I-50125 Firenze, Italy}


\date{\today}

\begin{abstract}
In the last years, a relationship has been established between the quantum Fisher information (QFI) and quantum entanglement. In the case of two-qubit systems, all pure entangled states can be made useful for sub-shot-noise interferometry while their QFI meets a necessary and sufficient condition \cite{PhysRevA.82.012337}. In $M$-qubit systems, the QFI provides just a sufficient condition in the task of detecting the degree of entanglement of a generic state \cite{Pezze_Smerzi_PRL102}. In our work, we show analytically that, for a large class of one-parameter non-optimal two-qubit states, the maximally entangled states are associated with stationary points of the QFI, as a function of such parameter.  We show, via numerical simulations, that this scenario is maintained for the generalisation of this class of states to a generic $M$-qubit system. Furthermore, we suggest a scheme for an interferometer able to detect the entanglement in a large class of two-spin states.

\end{abstract}

\maketitle


\section{Introduction}
\label{intro}

Entanglement is considered an essential resource for developing quantum-based technologies. It plays a fundamental role in quantum cryptography, quantum computation, teleportation, and in metrology based on the quantum phase estimation \cite{RevModPhys.81.865}. Despite its key role entanglement is yet elusive and the problem of its characterisation and quantification is still open \cite{PhysRevA.95.062116,PhysRevA.67.022320}. In the present article, we consider states depending on a parameter and we show that the maximally-entangled states are associated with stationary points of the quantum Fisher information (QFI) as a function of such parameter. Only in the case of optimal-states, the stationary points are indeed maxima for the quantum Fisher. With optimal-states, we mean states that maximise the value of the QFI via local unitary transformations.

In the present work, we investigate how to detect
entanglement by means of the value of the QFI even
in the case of non-optimised states. In particular, we
consider a class of entangled states derived as follows.
A system of $M$-qubits, initially in a separable state,
are entangled by means of the action of a non-local
unitary operator ${U}_0(\phi)$ that depends on a
continuous parameter $\phi$.
The entanglement degree of this state depends on
the value of such a parameter.
Starting from this entangled state, we generate
a class of trial states, by means of the action of
unitary local operators.
Such operations do not affect the entanglement
of these states \cite{PhysRevLett.80.2245}.
Then, we compute the QFI
associated with a local operator ${H}_1$ for these states.
\emph{We found that the maximally entangled states are
associated with stationary points of the quantum Fisher
information, as a function of $\phi$}.
In particular, the maximally entangled states
do not always correspond to maxima of the QFI, in fact,
we report an explicit example where they
are associated with a local minimum of the QFI.
\emph{Furthermore, for the case $M=2$ we investigate in
detail the symmetry
properties under spin exchange of the entangled states.
We relate the behaviour of the QFI
and the Husimi function to the breaking of such symmetry
that is induced by $U_0(\phi)$.
Finally, we suggest a scheme for an interferometer
able to detect the entanglement in a large class of
two-spin states.}

The paper is structured as follows.
In Sec. \ref{ent}, we derive the entangled states and summarise
their basic properties.
In Sec. \ref{squeez}, we highlight the relation between squeezing condition and entanglement condition in the case M=2.
In Sec. \ref{qfihf}, we define the trial states in the case M=2, calculate
either the QFI and the Husimi function of such states and we compare the results.
In Sec. \ref{results}, we discuss the link between the behaviour
of the QFI and the degree entanglement of these states.
In Sec. \ref{qfiM}, we show the numerical results of the QFI obtained in the case of $M>2$ qubits.
Finally, in Sec. \ref{conclusions}, we conclude with some remarks.

\section{Entangled states}
\label{ent}
The entanglement-generating operator we use is the
one introduced in Ref. \cite{briegel_PRL86_910}. 
We denote by $\sigma^j_x$, $\sigma^j_y$ and $\sigma^j_z$
the Pauli matrices operating on the $j$-th qubit ($j=1,\ldots , M$).
Furthermore, we denote with $\Pi^j_0=(\mathbb{I}+\sigma^j_z)/2$
and $\Pi^j_1=(\mathbb{I}-\sigma^j_z)/2$ the projector operators
onto the eigenstates of $\sigma^j_z$, $|0\rangle_j$ (with
eigenvalue $+1$)
and $|1\rangle_j$ (with eigenvalue $-1$),
respectively.
The entanglement-generating operator is
\begin{equation}
U_0(\phi) = \exp (-i \phi H_0) \, ,
\label{U0}
\end{equation}
where
\begin{equation}
H_0 = \sum^{M-1}_{j=1}\Pi^j_0 \Pi^{j+1}_1 \, .
\label{H0}
\end{equation}
Initially, the system is prepared in the state 
\begin{equation}
|r, 0 \rangle = \bigotimes^{M-1}_{j=0}
\dfrac{1}{\sqrt{2}}(|0\rangle_j + |1\rangle_j)
  \, ,
\label{r0}
\end{equation}
that is a tensor product of $\sigma_x^j$\textsc{\char13}s eigenstates.
The action of the non-local unitary operator $U_0$ on the initial (separable)
state $|r,0\rangle$ gives the state vector
\begin{equation}
|r, \phi \rangle= U_0(\phi)|r, 0 \rangle  \, .
\label{state-phi-M}
\end{equation}
By varying $\phi$, the degree of entanglement of the state
$|r,\phi\rangle$ varies accordingly. 
In particular, the values $\phi = (2k+1)  \pi$, ($k\in \mathbb{Z}$)
give the maximally entangled state, see Ref. \cite{briegel_PRL86_910}. 

We are interested in establishing a simple way to detect the maximally entangled states even in the case of non-optimised states since the latter is the most likely condition in an experimental situation. In the following, we will report analytic calculations in the case of two qubits and numeric analysis of the general case of many qubits that confirms the scenario of the case $M=2$.

\section{$\mathbf{M=2:}$ squeezing properties}
\label{squeez}
In the case $M=2$ state \eqref{state-phi-M}   results
\begin{equation}
|r, \phi \rangle= 
\frac{1}{2} \left( |00\rangle + e^{-i\phi}|01\rangle + |10\rangle 
+ |11 \rangle \right) \, .
\label{state-phi}
\end{equation}
For $\phi =2 k  \pi$ with $k\in \mathbb{Z}$, this state
is separable, whereas for all the other choices of the
value $\phi$, it is entangled. The maximally entangled
state 
\begin{equation}
|r, (2k+1)  \pi \rangle= \dfrac{1}{2}(|00\rangle -|01\rangle + |10\rangle 
+ |11 \rangle) \, ,
\label{rpi} 
\end{equation}
up to a a local unitary transformation on qubit $2$, may be written in
the form of a Bell state.
The squeezing properties of state \eqref{state-phi} can be obtained from
variances and expectation values of the angular momentum operators on this
state.
In fact, according to the criterion of spin squeezing by 
Kitagawa and Ueda in Ref. \cite{PhysRevA.47.5138}, a state
of a spin-$J$ system is squeezed if there exists a direction
${\bf n}$, orthogonal to the mean total-spin
$\langle {\bf J} \rangle$, such that
\begin{equation}
\xi^2 = 2 (\Delta J_{\bf n})^2/J < 1
\label{ss-cr}
\end{equation}
where $J_{\bf n} = {\bf J} \cdot {\bf n}$.
The expectation values and the uncertainties for the 
components of the total
angular momentum operator
${\bf J} = ({\boldsymbol \sigma}^1 +
{\boldsymbol \sigma}^2)/2$ on the state  
\eqref{state-phi} result
\begin{equation}
\begin{aligned}
\nonumber
&\langle J_x \rangle = \frac{1+\cos(\phi)}{2} \, , 
\  (\Delta J_x)^2 = 
\frac{2 -\cos(\phi) -\cos^2(\phi)}{4}
\, ,
 \\ \nonumber
&\langle J_y\rangle = 0 \, , \quad
 (\Delta J_y)^2 = \frac{1+\cos(\phi)}{4} \, ,
\\ \nonumber
&\langle J_z\rangle = 0 \, , \quad
(\Delta J_z)^2  = \frac{1}{2} 
\, .
\end{aligned}
\end{equation}
Therefore, in the present case $\langle {\bf J} \rangle$ is oriented along the $x$-axis, by choosing ${\bf n}$ along the $y$-axis form Eq. \eqref{ss-cr} we get the inequality
\begin{equation}
\cos(\phi) < 1 \, .
\label{ss-cr-1}
\end{equation}
Hence, for $\phi \neq 2 \pi k$, ($k\in \mathbb{Z}$) the state \eqref{state-phi} is squeezed, and the condition of squeezing for state \eqref{state-phi} is equivalent to that of entanglement. Furthermore, the maximally entangled state \eqref{rpi}, for which $\phi = (2k+1)\pi$, ($k\in \mathbb{Z}$),  is also the maximally squeezed state since this state corresponds also to the configuration where the inequality \eqref{ss-cr-1} is better satisfied. In the next section, we will confirm this scenario also through the Husimi function associated with a class of states that contains state \eqref{state-phi}.

\section{$\mathbf{M=2:}$ Quantum Fisher information and Husimi function}
\label{qfihf}

A rather natural choice for the local operator to evaluate the quantum Fisher information is
\begin{equation}
H_1 = J_z =  \frac{\sigma_z^1 + \sigma_z^2}{2} \, .
\label{H1}
\end{equation}
A direct calculation of the QFI
on the state \eqref{state-phi} gives
\begin{equation}
{\cal F}_q(|r,\phi\rangle , H_1) = 4  (\Delta H_1)^2  = 2 \,.
\label{fisher_H1_rphi}
\end{equation}
Since in the last expression the parameter $\phi$ is missing, in this case, the value of the QFI does not distinguish between entangled and non-entangled states. This is not an issue since, for a pure two-qubit state, the condition
${\cal F}_q > 2$ is just sufficient for particle entanglement, whereas the limit for a separable state is ${\cal F}_q \leq 2$ \cite{Pezze_Smerzi_PRL102}.

In order to investigate the possibility to exploit QFI in the task of witnessing the entanglement, also in the case of non-optimal states, it is necessary to consider a more general class of trial states. Such class is derived by applying local unitary transformations on the state \eqref{state-phi}. The entanglement properties of the original state are unaltered. In particular, the trial state $|s(\phi,\varphi_1,\varphi_2)\rangle$ is achieved by applying two separated rotations around the $y$ axis to the spins $1$ and $2$. The first spin is rotated by an angle $\varphi_1$ and the second of $\varphi_2$, thus it reads
\begin{equation}
\begin{split}
|s(\phi,\varphi_1,\varphi_2)\rangle &=
e^{-i\frac{\varphi_1}{2} \sigma^1_y }
e^{-i\frac{\varphi_2}{2} \sigma^2_y} 
|r, \phi\rangle = \\
& \dfrac{1}{2}\sum^1_{k,n=0} a_{kn} |k\, n\rangle\, ,
\label{try}
\end{split}
\end{equation}
where
\begin{equation}
\begin{aligned}
a_{00} &= \left( c_2 -s_2 e^{-i\phi} \right) c_1 -
\left( c_2 -s_2 \right) s_1 \, , \\
a_{01} &= \left( c_2 e^{-i\phi} + s_2  \right) c_1 -
\left( c_2 + s_2 \right) s_1 \, , \\
a_{10} &= \left( c_2 - s_2  \right) c_1 +
\left( c_2 - s_2 e^{-i\phi} \right) s_1 \, , \\
a_{11} &= \left( c_2 + s_2  \right) c_1 +
\left( c_2 e^{-i\phi} + s_2  \right) s_1 \, ,
\end{aligned}
\end{equation}
and $c_j =\cos(\varphi_j/2)$, $s_j =\sin(\varphi_j/2)$, $j=1,2$. 

\begin{figure}[ht] 
\begin{center}
{ 
\includegraphics[width=\linewidth]{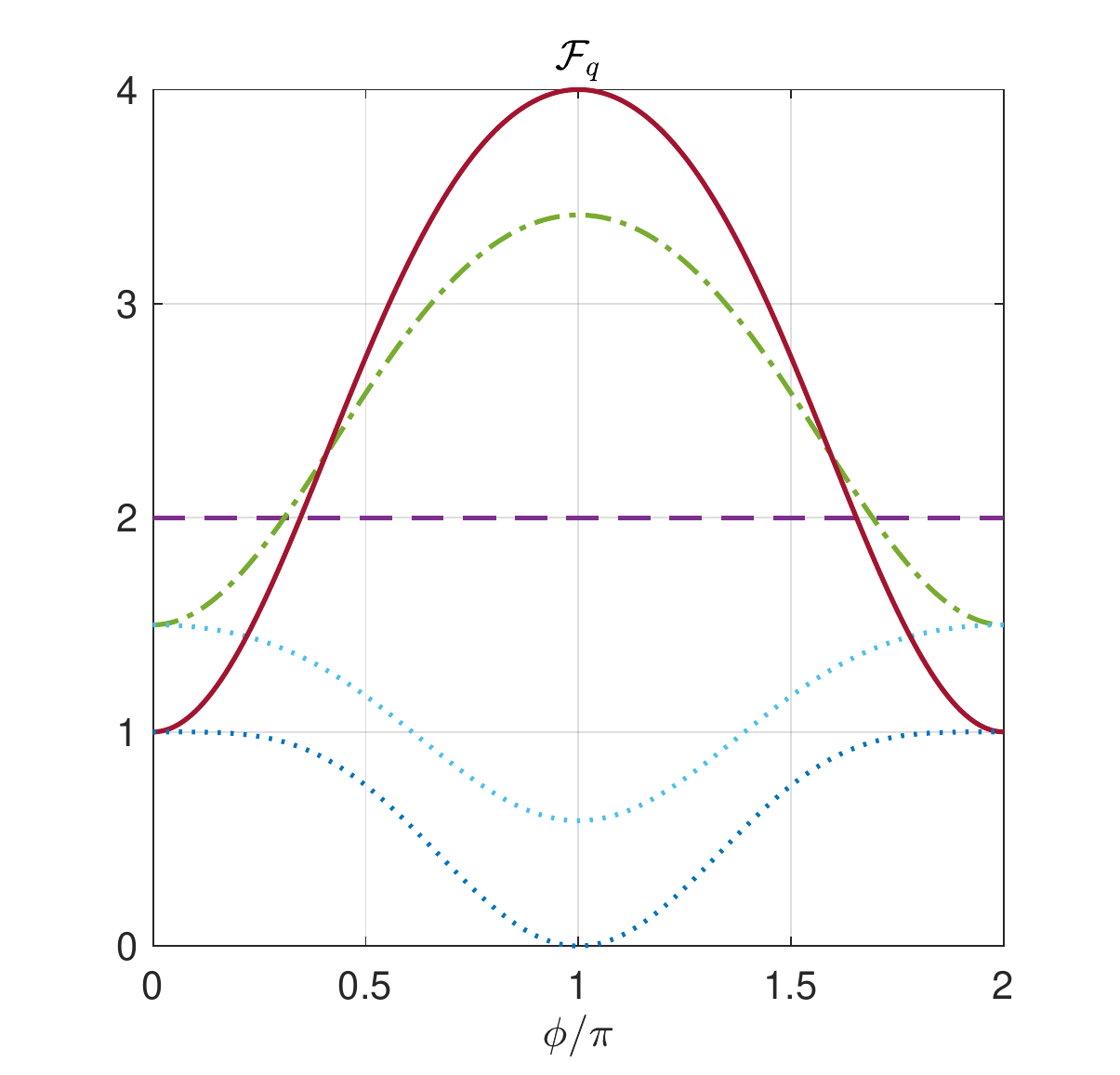}
}
\end{center}
\caption{The figure shows ${\cal F}_q(|s\rangle , H_1)$ vs $\phi$ for several choices of  $(\varphi_1,\varphi_2)$. In dashed line we report the plot corresponding to the choices $(0,0), (0,\pi), (\pi,0), (\pi,\pi)$ for which the QFI results constant. In continuous line the case corresponding to $(-\pi/2,0)$ while in dot-dashed line the choice $(-\pi/4,0)$. The upper dotted line represents the choice $(\pi/4,0)$ while the lower dotted line corresponds to the choice $(\pi/2,0)$.
}
\label{Fig01}
\end{figure}

In the following, we will refer to this state simply as $|s\rangle$ for brevity. The quantum Fisher information for this state results
\begin{equation}
\begin{aligned}
{\cal F}_q(|s\rangle , H_1) &= 2 + \sin(\varphi_2 − \varphi_1 ) \big[1 − \cos(\phi)\big] + \\[7pt]
& + \sin(\varphi_1 ) \sin(\varphi_2 ) \big[1 + \cos(\phi)\big] + \\[7pt]
& −\dfrac{1}{4} \big[1 + \cos(\phi)\big]^2 \big[\sin(\varphi_1 ) + \sin(\varphi_2)\big]^2 \,.
\label{fisher_H1_s}
\end{aligned}
\end{equation}
In this case, the QFI depends on the parameter that drives entanglement of the state $|s\rangle$. We note that \eqref{fisher_H1_s} is invariant under the transformations $(\varphi_1,\varphi_2) \to (-\varphi_2,-\varphi_1)$ and $\phi \to - \phi$. In Fig. \ref{Fig01} we plot ${\cal F}_q(|s\rangle , H_1)$ versus $\phi$ for several choices of the values of $\varphi_1$ and $\varphi_2$ as listed in the caption.
The dashed line in Fig. \ref{Fig01}, reports the plot in the cases $\varphi_1, \varphi_2 = 0,\pi$, for which ${\cal F}_q(|s\rangle , H_1)=2$. In fact, the action of such local transformations on \eqref{try}, at most, change the state \eqref{state-phi} for a sign factor.


From Eq. \eqref{fisher_H1_s} one deduces that the
detection of the maximally entangled state (which
corresponds to $\phi = \pi$) via stationary points of the QFI
is optimised, for instance, with the
values $\varphi_1 = - \pi/2$ and $\varphi_2 = 0$.
State \eqref{try} corresponding to this parameter choice thus represents an optimal
state.
However, in the case with $\varphi_1 = \pi/2$ and
$\varphi_2 = 0$, the maximally entangled state is associated
with a null value quantum Fisher information and thus exemplifies a 
non-optimal state (see lower dotted line in
Fig. \ref{Fig01}).
\emph{Although for two-qubit all pure entangled states can be
made useful for sub-shot-noise interferometry
\cite{PhysRevA.82.012337}, with non-optimal two-qubit
states the value of QFI provides a sufficient condition
in the task of detecting the degree of entanglement
of a generic state.}
In fact, in Ref. \cite{PhysRevA.82.012337} is shown that an entangled $2$-qubit
state can ever be optimised by local unitary transformations
in a way that when used as a probe in a linear interferometer
in order to estimate a phase, the phase sensitivity surpasses
the shot-noise limit $1/\sqrt{N}$, where $N=2$ is the number
of modes. In the case of 2 qubits, such limit is surpassed if
${\cal F}_q >2$.

A further state representation, useful for what we are going to discuss in the following, is given by the Husimi function. For a given state $|s(\phi,\varphi_1,\varphi_2)\rangle$, this is achieved by plotting onto the Bloch sphere the Husimi function \cite{Lee_PRA30_3308}
\begin{equation}
Q(\theta,\xi) = | \langle \theta ,\xi | 
s(\phi,\varphi_1,\varphi_2)\rangle|^2 \, ,
\label{husimi}
\end{equation}
where the coherent spin state 
\begin{equation}
\begin{aligned}
|\theta,\xi\rangle=\sum^M_{k=0}\cos^{k}\left(\dfrac{\theta}{2}\right)&\sin^{M-k}\left(\dfrac{\theta}{2} \right) \cdot \\
& \cdot e^{-i\xi (M-k)}\binom{M}{k}^{1/2}|M,k\rangle
\label{css}
\end{aligned}
\end{equation}
is given in terms of the Dicke states
\begin{equation}
    |M,k\rangle =  \binom{M}{k}^{-1/2} 
    \sum {\cal P}\{ |0\rangle^{\otimes k} \otimes |1\rangle^{\otimes M-k} \} \, .
\end{equation}
These are states completely symmetric under particle exchanges, indeed, they
are give as a sum  $\sum {\cal P}\{ \cdots\}$ on all the possible particle permutations. In the present case the number of particles considered is $M=2$ and it results
\begin{equation}
    |2,0\rangle = |11\rangle \, , \ 
    |2,1\rangle = (|01\rangle + |10\rangle)/\sqrt{2} \, , \
    |2,2\rangle = |00\rangle \, .
\end{equation}
\begin{figure}[ht] 
    \begin{center}
    {
        \includegraphics[width=1\linewidth]{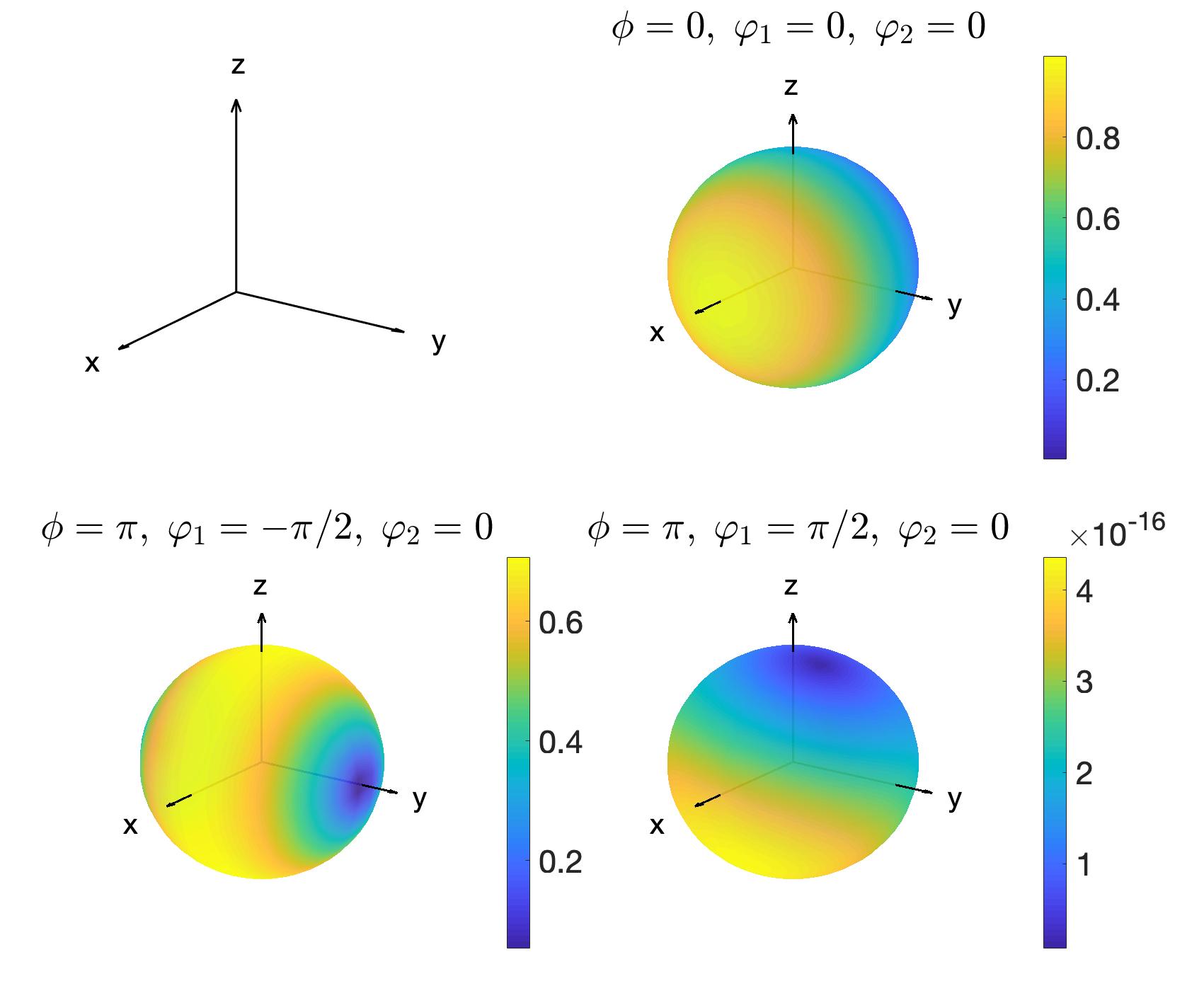}
    }
    \end{center}
    \caption{In the panel a) we plot $Q(\theta,\xi)$ for the separable state obtained with the choice $\phi=0$, $\varphi_1=0$ and $\varphi_2=0$. The panel b) reports the plot of $Q(\theta,\xi)$ for the maximally entangled state corresponding to $\phi=\pi$, $\varphi_1=-\pi/2$ and $\varphi_2=0$. The panel c) shows the maximally entangled state given by $\phi=\pi$, $\varphi_1=\pi/2$ and $\varphi_2=0$.}
    \label{Fig02}
\end{figure}

\section{Results}
\label{results}

\begin{figure}[ht]
    \begin{center}
    { 
        \includegraphics[width=1\linewidth]{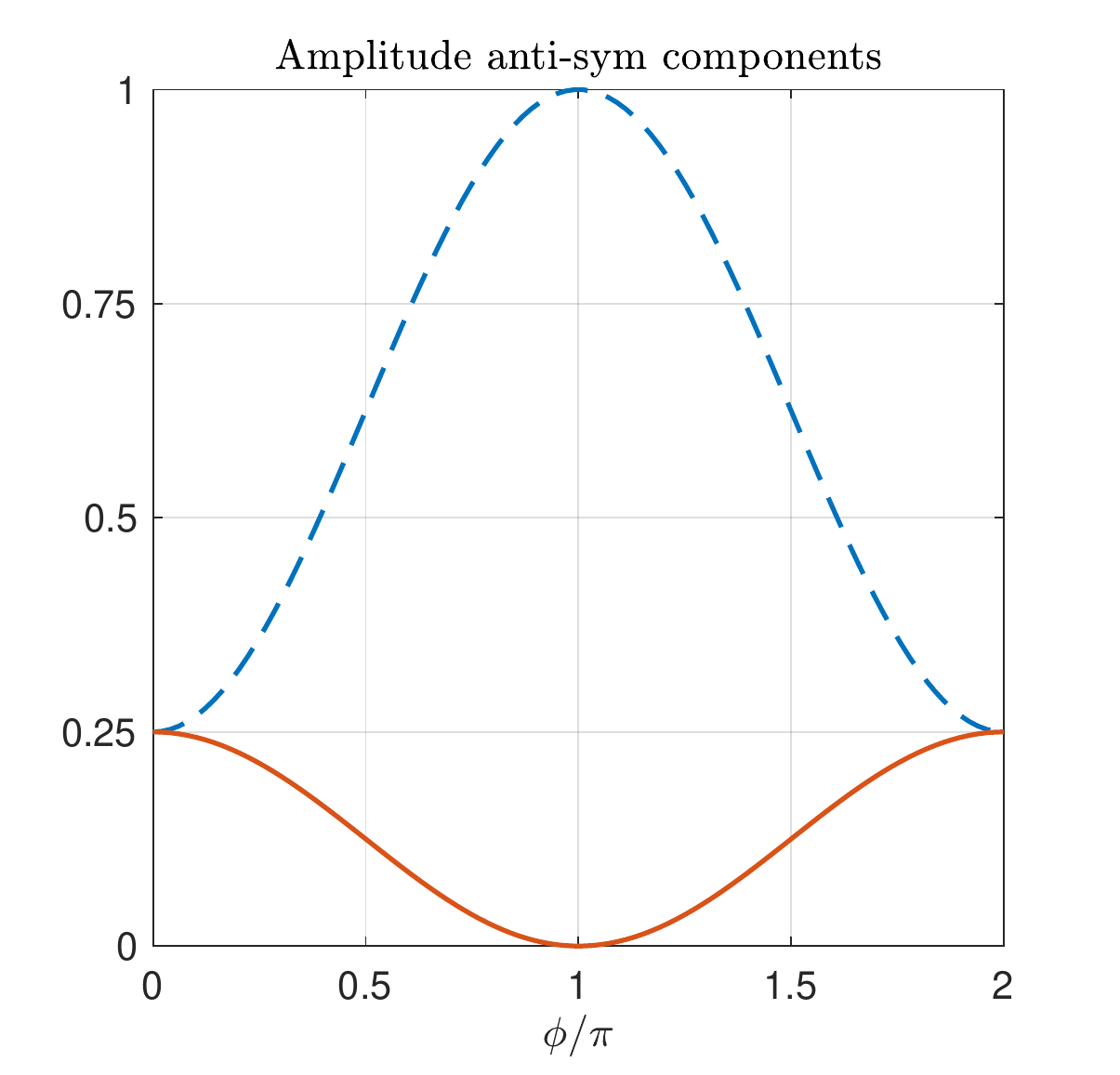}
    }
    \end{center}
    \caption{
    The figure plots the norm of the anti-symmetric component of the state $|s(\phi,\varphi_1,\varphi_2)\rangle$ for two different choices of rotations. In dashed line we report the plot corresponding to the choice $\varphi_1=\pi/2$ and $\varphi_2=0$, while the continuous line represents the anti-symmetric component vs $\phi$ corresponding to the choice $\varphi_1=-\pi/2$ and $\varphi_2=0$.
    }
    \label{Fig03}
\end{figure}

In this section, we compare the analysis of the entanglement properties of the state $|s\rangle$, achieved through the Husimi function and the Fisher information, in two distinctive cases: the first corresponds to the choice $\varphi_1=-\pi/2$, $\varphi_2=0$, and the second to the choice $\varphi_1=\pi/2$, $\varphi_2=0$. In both cases, we have analysed the Husimi function and the Fisher information as functions of the parameter $\phi$. Fig. \ref{Fig01} shows that in the case of the first choice ($\varphi_1=-\pi/2$, $\varphi_2=0$) for $|s\rangle$, the QFI has a peak in correspondence of the state with the maximum entanglement ($\phi=\pi$). Even the Husimi function detects the state with the maximum entanglement, in fact, Fig. \ref{Fig02} shows an increasing squeezing culminating in $\phi=\pi$ (see Fig. \ref{Fig02}b). In the case of the second choice ($\varphi_1=\pi/2$, $\varphi_2=0$), while the QFI has a minimum in correspondence of $\phi=\pi$ (see Fig. \ref{Fig01}), the Husimi function experiences a complete loss of information. This is shown in Fig. \ref{Fig02}c where such loss is evident and where no localisation on the Bloch sphere is observed. In both these examples, the state with the maximum degree of entanglement ($\phi = \pi$) is located at the zeros of the $d{\cal F}_q/d\phi$, thus at the stationary points. On the contrary, while with the first choice of rotations, the Husimi function is able to detect the maximally entangled state via its squeezing, in the second example, it is not able to determine such a state. Now, we take into account the symmetry under spin exchange of the two trial states in order to explain the poor performance of the Husimi function at witnessing entanglement. In this perspective, we split up the state $|s \rangle$ as a sum of its components: the symmetric $ |+\rangle = (|s \rangle + {\cal P}|s\rangle)/2$ and the antisymmetric $|-\rangle = (|s \rangle - {\cal P}|s\rangle)/2$. The Husimi function selects only the symmetric component of the state. Moreover, in case of the second couple of rotations, $\varphi_1 = \pi/2$,  $\varphi_2 =0$, the norm of $|+\rangle$ goes to zero for $\phi \to \pi$ and this explains the behaviour of the Husimi function highlighted above. A similar drawback does not occur with the QFI. The phenomenon discussed above is described in Fig. \ref{Fig03} where we plot the norm squared of the anti-symmetric component of the trial vectors versus $\phi$ in case of the two previous choices $\varphi_1=-\pi/2$, $\varphi_2=0$ and $\varphi_1=\pi/2$, $\varphi_2=0$. In the latter case, the norm-square of the anti-symmetric component $|-\rangle$ goes to $1$ at $\phi=\pi$.
Remarkably, the QFI is a faithful witness of entanglement for the whole class of states defined in Eq. \eqref{try}. In fact, from Eq. \eqref{fisher_H1_s} one obtains for the derivatives of the QFI with respect to $\phi$
\begin{equation}
    \begin{aligned}
        \dfrac{d {\cal F}_q  (|s\rangle , H_1)}{d\phi} = 
        4 \bigg\{
        \sin(\varphi_2 − \varphi_1 ) -
         \sin(\varphi_1 ) \sin(\varphi_2 ) + \\
         −\dfrac{1}{2} [\sin(\varphi_1 ) + \sin(\varphi_2)]^2
         [1 + \cos(\phi)]
         \bigg\}\sin(\phi) \, ,
        \label{dFQI01}
        \end{aligned}
\end{equation}
and this expression shows that \emph{the vanishing of $d{\cal F}_q/d\phi$ in correspondence of the maximally entangled state $(\phi=\pi)$ is a necessary condition}. In fact, even when $\sin(\phi) \neq 0$, it is possible to find values of $\varphi_1$ and $\varphi_2$ for which the derivative of the QFI is zero. Nevertheless, these stationary points correspond to non-optimised states that make useless the QFI. Examples of these non-optimised state are those corresponding to the dashed line of Fig. \ref{Fig01}.

\emph{Remarkably, the entanglement of these states is strongly linked to the symmetry breaking under spin exchange, that is induced by the unitary operator $U_0$.} This fact can be illustrated by the conceptual model of an interferometer able to detect the maximally entangled state, that we describe in the following. Such interferometer is obtained by the interference of two states. The first is derived from the initial state \eqref{state-phi} after the action of the following operator
\begin{equation}
    {\cal O}_1 = R_y(-\pi/2) R^1_y(\pi/2)\, ,
    \label{eq:O1}
\end{equation}
where $R^1_y(\pi/2)= e^{-i \sigma^1_y \pi/4}$
rotates by $\pi/2$ the first spin around the $y$-axis and
$ R_y(-\pi/2) =e^{i (\sigma^1_y + \sigma^2_y)\pi/4} $
rotates by $-\pi/2$ both the spins around the $z$-axis.
The second state is obtained starting from the same initial state, under the action of the operator
\begin{equation}
{\cal O}_2 =  {\cal P} R_y(-\pi/2) R^2_y(\pi/2)\, ,
\label{eq:O2}
\end{equation}
where $R^2_y(\pi/2)= e^{-i \sigma^2_y \pi/4}$
rotates by $\pi/2$ the
second spin around the $y$-axis.
Thus, the interference between the two states
\begin{equation}
{\cal A} (\phi)= \langle r,\phi |{\cal O}^\dagger_2 {\cal O}_1|r,\phi \rangle
= \dfrac{1}{2} (1+ \cos (\phi)) 
\label{calA}
\end{equation}
is completely destructive in the case of the maximally entangled state.
The link between the entanglement of these states and the symmetry
breaking induced by $U_0(\phi)$ is highlighted by recalling the
definition of state \eqref{state-phi}, $|r,\phi\rangle =
U_0(\phi)|r,0\rangle$  and noting that the state \eqref{r0}
is fully symmetric under spin exchange, that is ${\cal P}
|r,0\rangle = |r,0\rangle$. By using these relations in
Eq. \eqref{calA} one easily gets 
\begin{equation}
{\cal A} (\phi)= \langle r,0 |U^\dagger_0(\phi) {\cal P} U_0(\phi)
| r,0 \rangle \, .
\label{high}
\end{equation}
The latter relation emphasises that the operator $U_0(\phi)$
in addition to generating the entanglement it is responsible
for the ${\cal P}$-symmetry breaking.

\begin{figure}[ht]
\begin{center}
{ 
\includegraphics[width=1\linewidth]{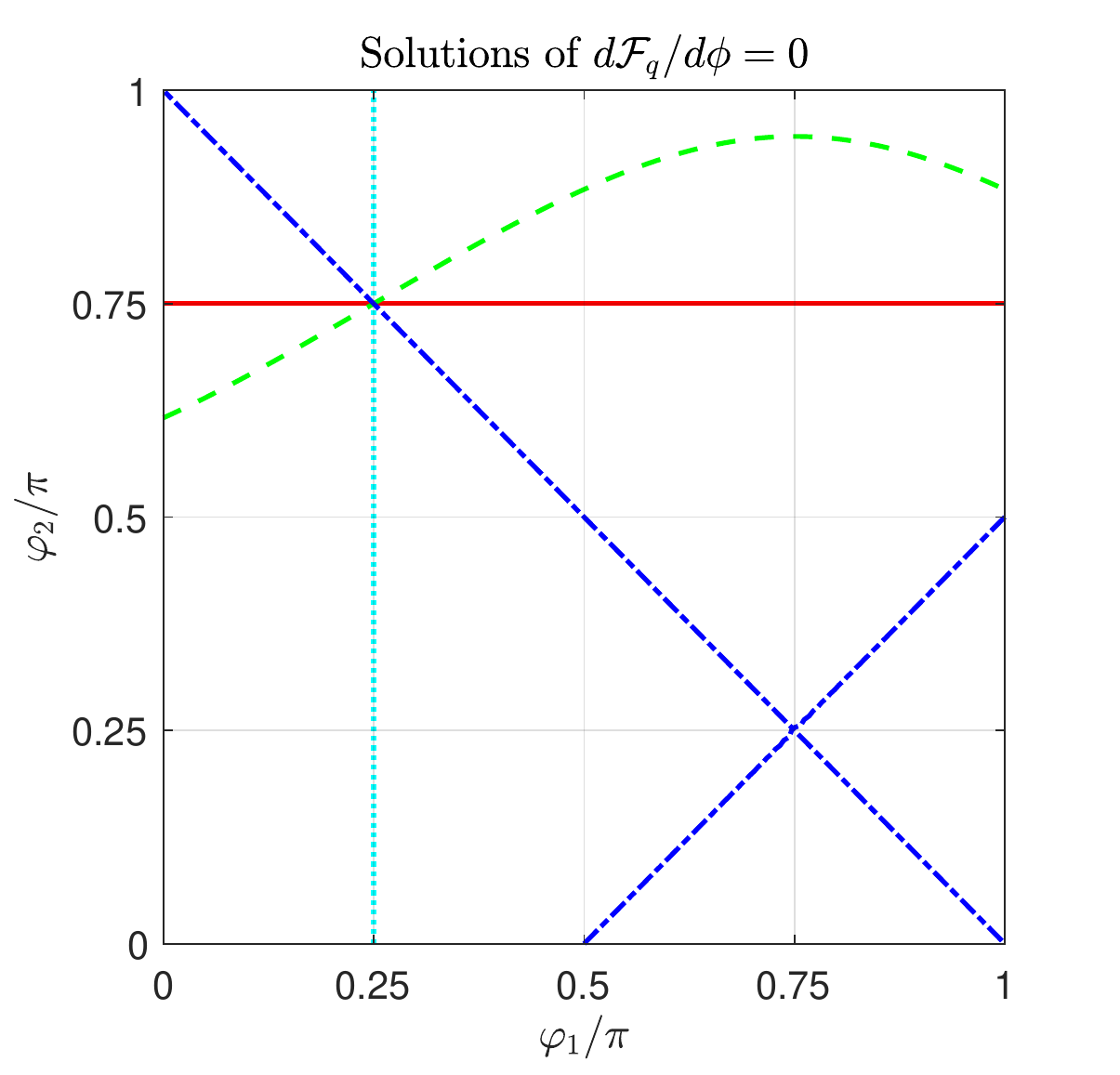}
}
\end{center}
\caption{
The figure shows the solutions $(\varphi_1, \varphi_2)$ for
$d{\cal F}_q/d\phi=0$, evaluated with $\phi=\pi$ and $\theta_1=\pi/3$,
and for some values of $\theta_2$.
The continuous line corresponds to the choice of $\theta_2=0$.
The dot-dashed one corresponds to $\theta_2=\pi/3$.
With the dotted line, we report the solutions corresponding to $\theta_2=\pi/2$,
and in dashed line the case corresponding to $\theta_2=(3/4)\pi$.
It is noteworthy the presence of the quadruple point in
$\varphi_1=\pi/4$ and $\varphi_2=3\pi/4$ that corresponds to the second
of the special solutions \eqref{magicnumbers}.
}
\label{Fig04}
\end{figure}
The class of states of Eq. \eqref{try} can be further enlarged by adding a
rotation around the $x$-axis for each spin, in the following way
\begin{equation}
\begin{aligned}
|s(\phi, & \theta_1,\theta_2,\varphi_1,\varphi_2)\rangle = \\[10pt] 
& = e^{-i\frac{\theta_1}{2} \sigma^1_x} e^{-i\frac{\theta_2}{2} \sigma^2_x}
e^{-i\frac{\varphi_1}{2} \sigma^1_y} e^{-i\frac{\varphi_2}{2} \sigma^2_y}
|r, \phi\rangle \, .
\label{try_general}
\end{aligned}
\end{equation}
By direct calculations, it is possible to derive the QFI, its
derivative with respect to $\phi$  and then evaluating the
latter at $\phi=\pi$. 
For brevity, we report just the latter result,
\begin{equation}
\begin{aligned}
\left. \frac{d{\cal F}_q}{d\phi} \right|_{\phi=\pi} &=\cos{\theta_1}\sin{\theta_2}(\sin{\varphi_1}-\cos{\varphi_1})+ \\
&-\sin{\theta_1}\cos{\theta_2}(\sin{\varphi_2}+\cos{\varphi_2}) \, .
\label{dv_phi}
\end{aligned}
\end{equation}
Let us determine the zeros for Eq.  \eqref{dv_phi}.
This results into solving the following expression 
\begin{equation}
\begin{aligned}
\tan{\theta_2}\sin\left(\varphi_1-\frac{\pi}{4}\right) = \tan{\theta_1}\sin\left(\varphi_2+\frac{\pi}{4}\right) \, .
\label{dv_phi_equal_0}
\end{aligned}
\end{equation}
Among all the infinite solutions, we have identified four couples
of \emph{magic numbers},
\begin{equation}
\begin{aligned}
\left(\theta_1=0,\theta_2=0 \right) \, , \quad
\left(\displaystyle\varphi_1=\frac{\pi}{4} ,
\displaystyle\varphi_2=\frac{3}{4}\pi 
\right) 
\, , \\
\left(\theta_1=0 ,
\displaystyle\varphi_1=\frac{\pi}{4}
\right) \, , \quad
\left(
\theta_2=0 ,
\displaystyle\varphi_2=\frac{3}{4}\pi
\right) \, .
\label{magicnumbers}
\end{aligned} 
\end{equation}
In fact, by choosing one of these couples
Eq. \eqref{dv_phi_equal_0} is satisfied independently
from the value assigned to the remaining angles.
In addition to these special solutions of the Eq. \eqref{dv_phi_equal_0},
additional infinite solutions exist.

\section{$\mathbf{M>2:}$ Quantum Fisher information }
\label{qfiM}
\begin{figure}[h]
    \begin{center}
    { 
        \includegraphics[width=1\linewidth]{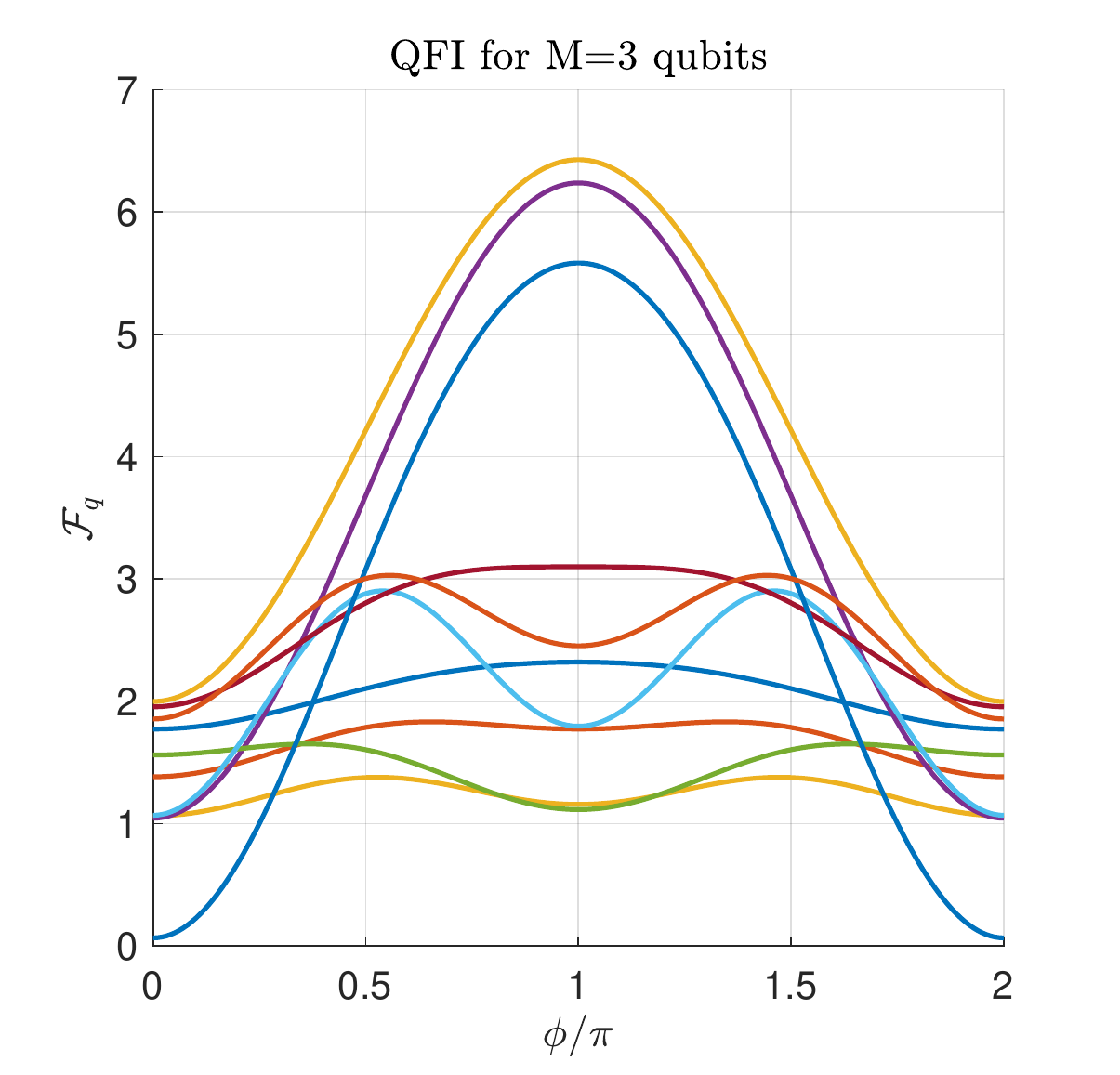}
    }
    \end{center}
    \caption{The figure plots  ${\cal F}_q(|s(\phi,{\bm \varphi})\rangle , H_1)$ vs $\phi$ for a system dimension $M=3$ and several randomly chosen values of the angles $\varphi_j$.}
    \label{FigM3}
\end{figure}
Just as an example, we have plotted in Fig. \ref{Fig04} some solutions ($\varphi_1$, $\varphi_2$)
corresponding to the choices $\theta_2 = 0,\pi/3,\pi/2,3\pi/4$ and maintaining the fixed value $\theta_1 = \pi/3$. In the figure is evident the presence of the quadruple point corresponding to the second of the special solutions just discussed.
\begin{figure}[h]
    \begin{center}
    { 
    \includegraphics[width=1\linewidth]{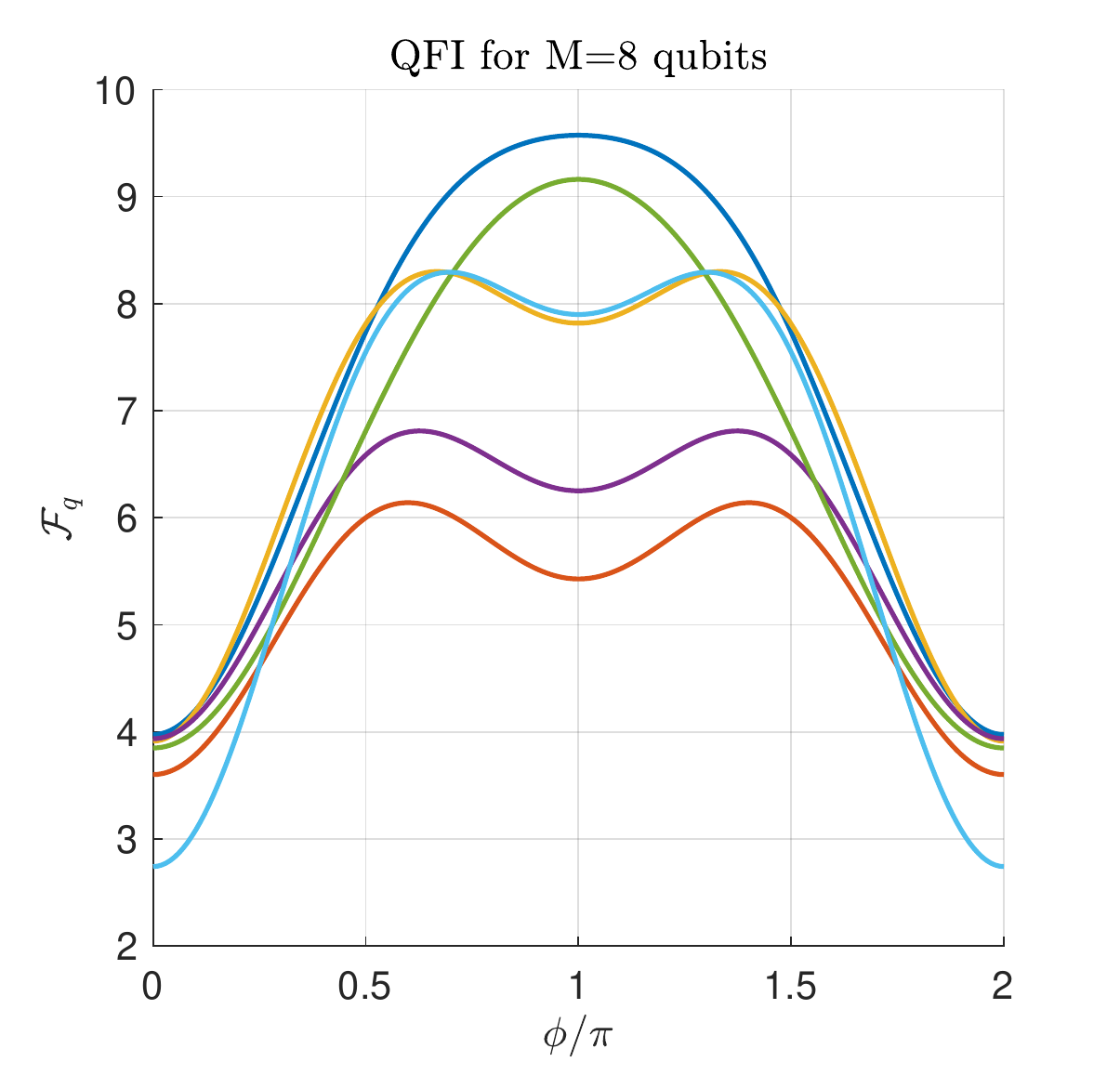}
    }
    \end{center}
    \caption{The figure plots  ${\cal F}_q(|s(\phi,{\bm \varphi})\rangle , H_1)$ vs $\phi$ for a system dimension $M=8$ and several randomly chosen values of the angles $\varphi_j$.}
    \label{FigM8}
\end{figure}
In the general case $M>2$, state \eqref{state-phi-M} results
\begin{equation}
\begin{aligned}
|r, \phi \rangle &= U_0(\phi)|r, 0 \rangle = \\ 
& 2^{-M/2} \sum^{2^M-1}_{k=0} 
\sum^{n(k)}_{j=0} \binom{n(k)}{j} \alpha^j
|k\rangle
 \, ,
\end{aligned}
\end{equation}
where 
\begin{equation}
\alpha = (e^{-i\phi} -1) \, ,
\end{equation}
and the kets $|k \rangle$, for $k=0,\ldots,2^M-1$ are
\[
|0\rangle =
|0 \cdots  0 \rangle \, , \, \,
|1\rangle =
|0 \cdots 0 1 \rangle
, \ldots ,
|2^M-1\rangle =
|1 \cdots 1 \rangle \, .
\]
For $\phi =2 k  \pi$ with $k\in \mathbb{Z}$, this state is separable, whereas for all the other choices of the value $\phi$, it is entangled. The maximally entangled states correspond to $\phi=(2k+1)\pi$ with $k\in \mathbb{Z}$ \cite{briegel_PRL86_910}. The trial state $|s(\phi,{\bm \varphi})\rangle$ is achieved by applying distinct rotations to each spin around the $y$ axis. The $j$-th spin is rotated by an angle $\varphi_j$, $j=1,\ldots,M$, thus it results
\begin{equation}
|s(\phi,{\bm \varphi})\rangle =\prod_j
e^{-i\frac{\varphi_j}{2} \sigma^j_y }
|r, \phi\rangle \, .
\label{tryM}
\end{equation}
As in the two-qubit case, we consider the operator 
\begin{equation}
    H_1 = J_z = \frac{1}{2} \sum^M_{j=1} \sigma^j_z \, ,
    \label{H1M}
\end{equation}
and we have evaluated the quantum Fisher ${\cal F}_q(|s(\phi,{\bm \varphi})\rangle , H_1)$ for several system dimension $M$ and several randomly chosen values of the angles $\varphi_j$.
\begin{figure}[h!]
    \begin{center}
    {
        \includegraphics[width=1\linewidth]{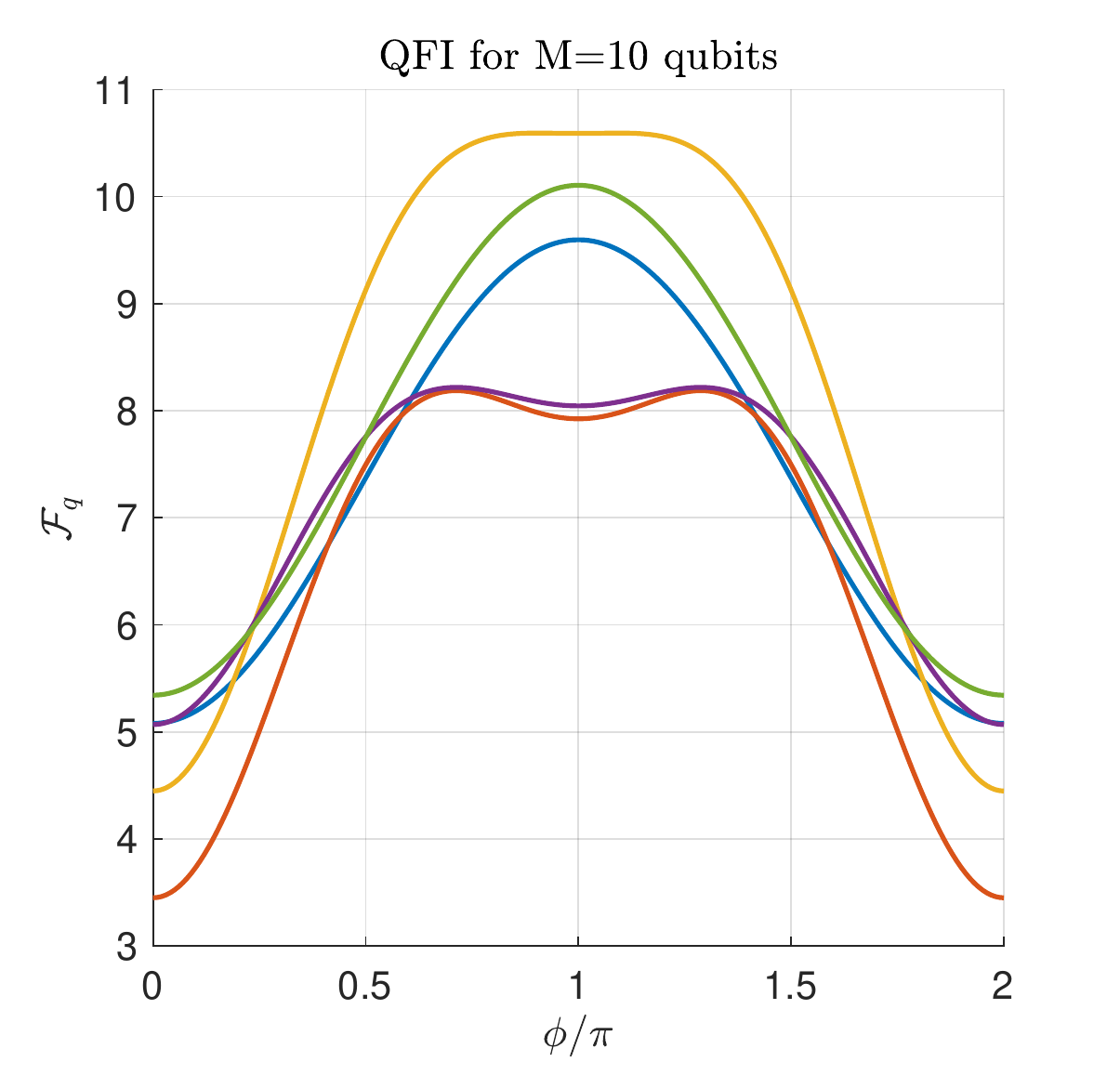}
    }
    \end{center}
    \caption{The figure plots  ${\cal F}_q(|s(\phi,{\bm \varphi})\rangle , H_1)$ vs $\phi$ for a system dimension $M=10$ and several randomly chosen values of the angles $\varphi_j$.}
    \label{FigM10}
\end{figure}
In Figs. \ref{FigM3},\ref{FigM8} and \ref{FigM10} we plot ${\cal F}_q(|s\rangle , H_1)$ versus $\phi$ for several choices of the values of $\varphi_1$ and $\varphi_2$. Fig. \ref{FigM3} refers to the case $M=3$, Fig. \ref{FigM8} to the case $M=8$ and Fig. \ref{FigM10} to the choice $M=10$. 
Figs. \ref{FigM3},\ref{FigM8} and \ref{FigM10}, show that in all the cases considered, the maximally entangled states, corresponding to $\phi=\pi$ \cite{briegel_PRL86_910,cocchiarella2019entanglement}, are associated with stationary points of the quantum Fisher information, as a function of $\phi$, thus confirming the result previously discussed for the case $M=2$.

\section{Conclusions}
\label{conclusions}
In summary, we have shown that for a large class of one-parameter M-qubit states, the maximally entangled states are associated with stationary points of the QFI, as a function of such parameter. Only in the case of optimal-states, these stationary points are maxima for the quantum Fisher. For the case $M=2$ we have also investigated the link between the breaking of the symmetry under the spin exchange and the entanglement induced by the unitary transformation $U_0(\phi)$. Finally, we have proposed a scheme for an interferometer that, exploiting such link, it is useful to detect the entanglement in a large class of two-spin states.

\begin{acknowledgments}
We are grateful to A. Smerzi, M. Gessner, M. Gabbrielli and L. Pezz\'e for useful discussions.\\

R. F. thanks the support by the QuantERA project “Q-Clocks” and the European Commission.
\end{acknowledgments}

\hfill
\bibliography{references}

\begin{thebibliography}{10}%
\makeatletter
\providecommand \@ifxundefined [1]{%
 \@ifx{#1\undefined}
}%
\providecommand \@ifnum [1]{%
 \ifnum #1\expandafter \@firstoftwo
 \else \expandafter \@secondoftwo
 \fi
}%
\providecommand \@ifx [1]{%
 \ifx #1\expandafter \@firstoftwo
 \else \expandafter \@secondoftwo
 \fi
}%
\providecommand \natexlab [1]{#1}%
\providecommand \enquote  [1]{``#1''}%
\providecommand \bibnamefont  [1]{#1}%
\providecommand \bibfnamefont [1]{#1}%
\providecommand \citenamefont [1]{#1}%
\providecommand \href@noop [0]{\@secondoftwo}%
\providecommand \href [0]{\begingroup \@sanitize@url \@href}%
\providecommand \@href[1]{\@@startlink{#1}\@@href}%
\providecommand \@@href[1]{\endgroup#1\@@endlink}%
\providecommand \@sanitize@url [0]{\catcode `\\12\catcode `\$12\catcode
  `\&12\catcode `\#12\catcode `\^12\catcode `\_12\catcode `\%12\relax}%
\providecommand \@@startlink[1]{}%
\providecommand \@@endlink[0]{}%
\providecommand \url  [0]{\begingroup\@sanitize@url \@url }%
\providecommand \@url [1]{\endgroup\@href {#1}{\urlprefix }}%
\providecommand \urlprefix  [0]{URL }%
\providecommand \Eprint [0]{\href }%
\providecommand \doibase [0]{http://dx.doi.org/}%
\providecommand \selectlanguage [0]{\@gobble}%
\providecommand \bibinfo  [0]{\@secondoftwo}%
\providecommand \bibfield  [0]{\@secondoftwo}%
\providecommand \translation [1]{[#1]}%
\providecommand \BibitemOpen [0]{}%
\providecommand \bibitemStop [0]{}%
\providecommand \bibitemNoStop [0]{.\EOS\space}%
\providecommand \EOS [0]{\spacefactor3000\relax}%
\providecommand \BibitemShut  [1]{\csname bibitem#1\endcsname}%
\let\auto@bib@innerbib\@empty
\bibitem [{\citenamefont {Hyllus}\ \emph {et~al.}(2010)\citenamefont {Hyllus},
  \citenamefont {G\"uhne},\ and\ \citenamefont {Smerzi}}]{PhysRevA.82.012337}%
  \BibitemOpen
  \bibfield  {author} {\bibinfo {author} {\bibfnamefont {P.}~\bibnamefont
  {Hyllus}}, \bibinfo {author} {\bibfnamefont {O.}~\bibnamefont {G\"uhne}}, \
  and\ \bibinfo {author} {\bibfnamefont {A.}~\bibnamefont {Smerzi}},\ }\href
  {\doibase 10.1103/PhysRevA.82.012337} {\bibfield  {journal} {\bibinfo
  {journal} {Phys. Rev. A}\ }\textbf {\bibinfo {volume} {82}},\ \bibinfo
  {pages} {012337} (\bibinfo {year} {2010})}\BibitemShut {NoStop}%
\bibitem [{\citenamefont {Pezz\'e}\ and\ \citenamefont
  {Smerzi}(2009)}]{Pezze_Smerzi_PRL102}%
  \BibitemOpen
  \bibfield  {author} {\bibinfo {author} {\bibfnamefont {L.}~\bibnamefont
  {Pezz\'e}}\ and\ \bibinfo {author} {\bibfnamefont {A.}~\bibnamefont
  {Smerzi}},\ }\href {\doibase 10.1103/PhysRevLett.102.100401} {\bibfield
  {journal} {\bibinfo  {journal} {Phys. Rev. Lett.}\ }\textbf {\bibinfo
  {volume} {102}},\ \bibinfo {pages} {100401} (\bibinfo {year}
  {2009})}\BibitemShut {NoStop}%
\bibitem [{\citenamefont {Horodecki}\ \emph {et~al.}(2009)\citenamefont
  {Horodecki}, \citenamefont {Horodecki}, \citenamefont {Horodecki},\ and\
  \citenamefont {Horodecki}}]{RevModPhys.81.865}%
  \BibitemOpen
  \bibfield  {author} {\bibinfo {author} {\bibfnamefont {R.}~\bibnamefont
  {Horodecki}}, \bibinfo {author} {\bibfnamefont {P.}~\bibnamefont
  {Horodecki}}, \bibinfo {author} {\bibfnamefont {M.}~\bibnamefont
  {Horodecki}}, \ and\ \bibinfo {author} {\bibfnamefont {K.}~\bibnamefont
  {Horodecki}},\ }\href {\doibase 10.1103/RevModPhys.81.865} {\bibfield
  {journal} {\bibinfo  {journal} {Rev. Mod. Phys.}\ }\textbf {\bibinfo {volume}
  {81}},\ \bibinfo {pages} {865} (\bibinfo {year} {2009})}\BibitemShut
  {NoStop}%
\bibitem [{\citenamefont {Sperling}\ and\ \citenamefont
  {Walmsley}(2017)}]{PhysRevA.95.062116}%
  \BibitemOpen
  \bibfield  {author} {\bibinfo {author} {\bibfnamefont {J.}~\bibnamefont
  {Sperling}}\ and\ \bibinfo {author} {\bibfnamefont {I.~A.}\ \bibnamefont
  {Walmsley}},\ }\href {\doibase 10.1103/PhysRevA.95.062116} {\bibfield
  {journal} {\bibinfo  {journal} {Phys. Rev. A}\ }\textbf {\bibinfo {volume}
  {95}},\ \bibinfo {pages} {062116} (\bibinfo {year} {2017})}\BibitemShut
  {NoStop}%
\bibitem [{\citenamefont {Giovannetti}\ \emph {et~al.}(2003)\citenamefont
  {Giovannetti}, \citenamefont {Mancini}, \citenamefont {Vitali},\ and\
  \citenamefont {Tombesi}}]{PhysRevA.67.022320}%
  \BibitemOpen
  \bibfield  {author} {\bibinfo {author} {\bibfnamefont {V.}~\bibnamefont
  {Giovannetti}}, \bibinfo {author} {\bibfnamefont {S.}~\bibnamefont
  {Mancini}}, \bibinfo {author} {\bibfnamefont {D.}~\bibnamefont {Vitali}}, \
  and\ \bibinfo {author} {\bibfnamefont {P.}~\bibnamefont {Tombesi}},\ }\href
  {\doibase 10.1103/PhysRevA.67.022320} {\bibfield  {journal} {\bibinfo
  {journal} {Phys. Rev. A}\ }\textbf {\bibinfo {volume} {67}},\ \bibinfo
  {pages} {022320} (\bibinfo {year} {2003})}\BibitemShut {NoStop}%
\bibitem [{\citenamefont {Wootters}(1998)}]{PhysRevLett.80.2245}%
  \BibitemOpen
  \bibfield  {author} {\bibinfo {author} {\bibfnamefont {W.~K.}\ \bibnamefont
  {Wootters}},\ }\href {\doibase 10.1103/PhysRevLett.80.2245} {\bibfield
  {journal} {\bibinfo  {journal} {Phys. Rev. Lett.}\ }\textbf {\bibinfo
  {volume} {80}},\ \bibinfo {pages} {2245} (\bibinfo {year}
  {1998})}\BibitemShut {NoStop}%
\bibitem [{\citenamefont {Briegel}\ and\ \citenamefont
  {Raussendorf}(2001)}]{briegel_PRL86_910}%
  \BibitemOpen
  \bibfield  {author} {\bibinfo {author} {\bibfnamefont {H.~J.}\ \bibnamefont
  {Briegel}}\ and\ \bibinfo {author} {\bibfnamefont {R.}~\bibnamefont
  {Raussendorf}},\ }\href {\doibase 10.1103/PhysRevLett.86.910} {\bibfield
  {journal} {\bibinfo  {journal} {Phys. Rev. Lett.}\ }\textbf {\bibinfo
  {volume} {86}},\ \bibinfo {pages} {910} (\bibinfo {year} {2001})}\BibitemShut
  {NoStop}%
\bibitem [{\citenamefont {Kitagawa}\ and\ \citenamefont
  {Ueda}(1993)}]{PhysRevA.47.5138}%
  \BibitemOpen
  \bibfield  {author} {\bibinfo {author} {\bibfnamefont {M.}~\bibnamefont
  {Kitagawa}}\ and\ \bibinfo {author} {\bibfnamefont {M.}~\bibnamefont
  {Ueda}},\ }\href {\doibase 10.1103/PhysRevA.47.5138} {\bibfield  {journal}
  {\bibinfo  {journal} {Phys. Rev. A}\ }\textbf {\bibinfo {volume} {47}},\
  \bibinfo {pages} {5138} (\bibinfo {year} {1993})}\BibitemShut {NoStop}%
\bibitem [{\citenamefont {Lee}(1984)}]{Lee_PRA30_3308}%
  \BibitemOpen
  \bibfield  {author} {\bibinfo {author} {\bibfnamefont {C.~T.}\ \bibnamefont
  {Lee}},\ }\href {\doibase 10.1103/PhysRevA.30.3308} {\bibfield  {journal}
  {\bibinfo  {journal} {Phys. Rev. A}\ }\textbf {\bibinfo {volume} {30}},\
  \bibinfo {pages} {3308} (\bibinfo {year} {1984})}\BibitemShut {NoStop}%
\bibitem [{\citenamefont {Cocchiarella}\ \emph {et~al.}(2019)\citenamefont
  {Cocchiarella}, \citenamefont {Scali}, \citenamefont {Ribisi}, \citenamefont
  {Nardi}, \citenamefont {Aissa},\ and\ \citenamefont
  {Franzosi}}]{cocchiarella2019entanglement}%
  \BibitemOpen
  \bibfield  {author} {\bibinfo {author} {\bibfnamefont {D.}~\bibnamefont
  {Cocchiarella}}, \bibinfo {author} {\bibfnamefont {S.}~\bibnamefont {Scali}},
  \bibinfo {author} {\bibfnamefont {S.}~\bibnamefont {Ribisi}}, \bibinfo
  {author} {\bibfnamefont {B.}~\bibnamefont {Nardi}}, \bibinfo {author}
  {\bibfnamefont {G.~B.~H.}\ \bibnamefont {Aissa}}, \ and\ \bibinfo {author}
  {\bibfnamefont {R.}~\bibnamefont {Franzosi}},\ }\href@noop {} {\bibfield
  {journal} {\bibinfo  {journal} {arXiv preprint arXiv:1908.03117}\ } (\bibinfo
  {year} {2019})}\BibitemShut {NoStop}%
\end{thebibliography}%

\end{document}